\documentclass[preprint,showpacs,amssymb,amsmath,superscriptaddress]{revtex4-1}
\usepackage{graphicx}
\usepackage{dcolumn}
\usepackage{bm} 
\usepackage{bbm}
\usepackage{ulem}
\usepackage{color}
\usepackage{epstopdf}
\usepackage{amssymb}
\usepackage{amsmath}
\usepackage{tikz}
\usetikzlibrary{arrows,calc}
\usepackage{comment}

\newcommand{\tr}{\text{Tr}} 
\newcommand{\A}{\mathcal{A}} 

\newcommand{\ket}[1]{\left| #1 \right>} 
\newcommand{\braket}[2]{\left< #1 \vphantom{#2} \right| \left. #2 \vphantom{#1} \right>} 

\begin{document}

\title{Topological protection of photonic mid-gap cavity modes}
\author{Jiho Noh*}
\affiliation{Department of Physics, The Pennsylvania State University, University Park, PA 16802, USA}
\author{Wladimir A. Benalcazar*}
\affiliation{Department of Physics, Institute for Condensed Matter Theory, University of Illinois at Urbana-Champaign, Illinois 61801, USA}
\author{Sheng Huang*}
\affiliation{Department of Electrical and Computer Engineering, University of Pittsburgh, Pittsburgh, Pennsylvania 15261, USA}
\author{Matthew J. Collins}
\affiliation{Department of Physics, The Pennsylvania State University, University Park, PA 16802, USA}
\author{Kevin Chen}
\affiliation{Department of Electrical and Computer Engineering, University of Pittsburgh, Pittsburgh, Pennsylvania 15261, USA}
\author{Taylor L. Hughes}
\affiliation{Department of Physics, Institute for Condensed Matter Theory, University of Illinois at Urbana-Champaign, Illinois 61801, USA}
\author{Mikael C. Rechtsman}
\affiliation{Department of Physics, The Pennsylvania State University, University Park, PA 16802, USA}

\date{\today}

\begin{abstract}
{\bf Defect modes in two-dimensional periodic photonic structures have found use in a highly diverse set of optical devices. For example, photonic crystal cavities confine optical modes to subwavelength volumes and can be used for Purcell enhancement of nonlinearity, lasing, and cavity quantum electrodynamics. Photonic crystal fiber defect cores allow for supercontinuum generation and endlessly-single-mode fibers with large cores.  However, these modes are notoriously fragile: small changes in the structure can lead to significant detuning of resonance frequency and mode volume.  Here, we show that a photonic topological crystalline insulator structure can be used to topologically protect the resonance frequency to be in the middle of the band gap, and therefore minimize the mode volume of a two-dimensional photonic defect mode.  We experimentally demonstrate this in a femtosecond-laser-written waveguide array, a geometry akin to a photonic crystal fiber.  The topological defect modes are determined by a topological invariant that protects zero-dimensional states (defect modes) embedded in a two-dimensional environment; a novel form of topological protection that has not been previously demonstrated.}

\end{abstract}

\maketitle

The field of topological photonics \cite{Lu2014} has as its central aim to topologically protect the flow of photons from the effects of parasitic scattering by the inevitable disorder that arises in device fabrication.  Photonic topological insulators (PTIs) \cite{Haldane2008, Soljacic2009, Carusotto2011,Hafezi2011,Fan2012,Rechtsman2013, Hafezi2013, Khanikaev2013, Wu2015} usually have edge states whereby photons travel along the edge of the structure in a robust way. These ideas were inherited from electronic materials in condensed matter physics, where this robustness was shown in the context of the two-dimensional quantum Hall \cite{Klitzing1980,Thouless1982} and quantum spin Hall \cite{Kane2005graphene,Kane2005z2,Bernevig2006,Molenkamp} insulators.  Two natural classes of PTIs are: (1) those that break time-reversal symmetry \cite{Soljacic2009, Rechtsman2013}, and achieve robustness because they have no counterpropagating partner at the same frequency on the same edge; or (2) those that preserve time-reversal symmetry, have a counterpropagating partner, but do not couple to it as long as the disorder respects certain symmetries \cite{Khanikaev2013, Hafezi2013, Wu2015}.  Only the first type achieves full robustness due to the total lack of an available state into which photons may scatter.  However, the ultimate goal of realizing an in-plane, time-reversal-broken optical topological insulator remains elusive due to the weakness of the Faraday effect at optical frequencies~\cite{Soljacic2009}, and the difficulty of achieving very fast modulation~\cite{LipsonModulator}.    

There are also one-dimensional topological systems that have end states that exist at the termination of a 1d lattice, and at frequencies in the center of the band gap.  These are zero-dimensional modes (i.e., localized in all directions), and therefore do not exhibit electronic transport.  The first of these was the Shockley state \cite{Shockley1939}, which has also been observed in photonics \cite{Malkova2009}; followed by solitons in the Su-Schreiffer-Heeger (SSH) dimerized chain \cite{Su1979}, which forms the basis for defect zero energy modes in the context of electronic systems, and Majorana modes in analogous superconducting wires \cite{Kitaev2001} (see other photonic manifestations in Refs. \cite{Weinstein2014, poli2015, slobozhanyuk2015, blanco2016}).  Any disorder introduced in these systems that respects the chiral or particle-hole symmetries \cite{Asboth,mondragon2014} will preserve the localized topological defect state pinned to the center of the gap.  It was recently shown \cite{TeoHughes2013, Benalcazar2014} that, surprisingly, even {{\it two}}-dimensional, time-reversal invariant topological crystalline insulator and superconductor structures can support zero-dimensional topological defect modes whose energies lie at mid-gap. This is a fundamentally different type of topological protection because, in the other cases, states being protected are a single dimension lower than the system dimension (i.e., one-dimensional edge states protected in a two-dimensional TI; or zero-dimensional end states protected in a one-dimensional system).  This leads to the natural question: can two-dimensional time-reversal invariant {\it photonic} insulators with crystalline symmetries be used to realize protected, maximally localized defect modes that must reside in the middle of the band gap with a stable frequency?  

If so, this would represent a novel mechanism for stabilizing defect modes in two-dimensional photonic crystal slabs \cite{painter1999two} and fibers \cite{russell2003photonic}.   Ordinary (non-topological) defect modes usually have frequencies that bifurcate from the band edges and do not naturally arise at the center of the gap; they are inherently sensitive to imperfections.  Indeed, if defects are too weak, their frequencies will lie close to the band edge from which they emerged; if they are too strong, their frequencies will cross the gap and lie too close to the opposite band edge.  Topological protection would provide more stable mode frequencies and tighter mode confinement (since mode size goes down with separation from the photonic band edge).  Even if perturbations arise that break the required symmetries (discussed below), the confined modes would {\it start out} at mid-gap rather than bifurcating from a band edge (hence requiring fine-tuning).  This would in turn enable more efficient coupling of photonic crystal cavity modes for scaling of quantum electrodynamics-based quantum information devices \cite{o2009photonic}; enhanced nonlinear optical effects and stronger coupling to quantum dots due to higher Purcell factors~\cite{englund2005controlling}; more efficient supercontinuum generation in fibers due to stronger confinement \cite{dudley2006supercontinuum} (or a higher yield in production with accompanying lower cost); among many other applications.  Moreover, since it would not require breaking time-reversal symmetry, it would mean experimental implementation can be straightforward in a number of different configurations.  This  mechanism is distinct from previous work on stabilized photonic cavity modes since it relies on, in-principle, loss-free dielectric structures (as opposed to epsilon-near-zero structures \cite{Engheta}), and rigorously pins the modes to be mid-gap (via a topological invariant) and not embedded within a band \cite{Mortessagne}.  

Here we present a model for the realization of such protected defect modes in two dimensions, and we experimentally demonstrate their presence. We use an array of evanescently-coupled waveguides written into fused silica by the femtosecond direct laser writing technique \cite{Szameit2005}; the geometry we employ was introduced in Ref. \cite{Benalcazar2014}, and explored theoretically in Ref. \cite{Wu2015}, although the latter does not predict the defect states that are our focus, and instead focused on the (gapped) edge states. The diffraction of light through this waveguide array is governed by the paraxial wave equation
\begin{equation}
i\partial_{z}\psi(\textbf{r},z)=-\frac{1}{2k_{0}}\nabla^{2}_{\textbf{r}}\psi(\textbf{r},z)-\frac{k_{0}\Delta n(\textbf{r})}{n_{0}}\psi(\textbf{r},z),
\label{eq:propagation}
\end{equation}
where $\psi(\textbf{r},z)$ is the envelope function of the electric field $\textbf{E}(\textbf{r},z)=\psi(\textbf{r},z)e^{i(k_{0}z-\omega t)}\hat{x}$, $k_{0}=2\pi n_{0}/\lambda$ is the wavenumber within the medium, $\lambda$ is the wavelength of light, $\nabla^{2}_{\textbf{r}}$ is the Laplacian in the transverse $(x,y)$ plane, and $\omega=2\pi c/\lambda$. Our medium is borosilicate glass with refractive index $n_{0}=1.5,$ and $\Delta n$ is the refractive index relative to $n_{0},$ which  acts as an effective potential in the analogous Schr{\"o}dinger equation \eqref{eq:propagation}. Assuming that light is tightly confined to the waveguides, we may employ the tight-binding approximation
\begin{equation}
i\partial_{z}\psi_{i}(z)=-\sum_{j}c_{ij}(\lambda)\psi_{j}(z),
\label{coupledmodeEq}
\end{equation}
where $\psi_{n}$ is amplitude in the $n$-th waveguide, and $c_{ij}$ is the coupling constant between waveguides $i$ and $j$.
Our photonic lattices are constant along the propagation direction $z,$ thus, we can explicitly write the $z$-dependence of the propagating modes in Eq. \ref{coupledmodeEq} as $\psi_n(z)=\psi_n e^{i \beta z}$. This leads to
\begin{equation}
\beta \psi_{i}=\sum_{j}c_{ij}(\lambda)\psi_{j},
\label{eigensystemEq}
\end{equation}
where $\beta$ plays the role of energy in the analogous Schr{\"o}dinger equation $H \psi_i = \beta \psi_i$, where $H_{ij}\equiv c_{ij}$.

In the transverse plane, the waveguide arrays in our photonic lattices have $C_{6}$ symmetry as shown in Fig. \ref{Fig1} \cite{Benalcazar2014, Wu2015}. The primitive lattice is triangular and each unit cell has six waveguides. Neighboring waveguides within the unit cell are separated by a distance $s$ (a parameter that we tune in the experiment), and the lattice constant is $L=50$ $\mu$m. The ratio between these two lengths $L$ and $s$ allows tuning between two topologically distinct gapped phases. For $L/s>3$ this structure is topologically trivial, at $L/s=3$ it is gapless, and for $L/s<3$, it is topologically non-trivial. Microscope images of experimental samples used in each of these three cases are shown in Figs. \ref{Fig1}a-c., and the corresponding tight-binding diagrams in Fig. \ref{Fig1}d-f. Notice that the lattice at the critical point $L/s=3$ is identical to photonic graphene (i.e., the perfect honeycomb lattice). 

As the usual hallmark of non-trivial topological phases is the existence of symmetry-protected states on the boundaries of the material, we calculate the energy $\beta$ spectrum (see Eq. \ref{eigensystemEq}) in a configuration with periodic boundaries in one direction but open in the other. In the trivial phase there are only bulk bands (Fig. \ref{Fig1}g), as expected. Starting from this trivial phase, the band gap decreases as $L/s$ approaches the critical value of $3$. At $L/s=3$ our structure is simply a photonic analogue of graphene with armchair edges, and its spectrum is gapless (and also without boundary modes) (Fig. \ref{Fig1}h). For values of $L/s<3$, the gap re-opens, but with additional  sub-gap bands, as shown in red in Fig. \ref{Fig1}i. Although the corresponding sub-gap states are localized at the edges, these states are gapped themselves and are not topologically protected. 
This differs from other topological phases with time-reversal (TR) symmetry, e.g., a quantum spin Hall (QSH) insulator, which exhibits gapless edge states protected by TR symmetry. The fundamental difference is that, while in spinfull electronic systems  TR leads to Kramers degeneracy, in our photonic system the time reversal operator $\hat{T}$ obeys $\hat{T}^2=+1$, and does not prevent the edge states from hybridizing and opening an energy gap. Indeed, TR-invariant photonic systems such as ours belong to class AI in the periodic 10-fold classification of topological phases \cite{Altland1997, Schnyder2009}, and in two dimensions this class does not exhibit non-trivial topological phases~\cite{Teo2010}. Thus, any non-trivial topological phenomena will have to necessarily arise from the existence of extra symmetries, e.g., discrete translation or point-group symmetries. This is the case in our model which has $C_6$ rotation symmetry and (approximate) chiral symmetry. These symmetries protect topological bound states on certain corners of the two-dimensional structure when in the topological phase, and pin their energy to $\beta = 0.$ 

While the $C_6$ symmetry of our lattice can readily be noticed, the existence of chiral symmetry is more subtle. In fact, the chiral symmetry only precisely exists when coupling between sites on the same sublattice (c.f., site colors in Fig. \ref{Fig1}d-f) are vanishing. This is a good approximation for our system as the coupling terms $c_{ij}(\lambda)$ decrease exponentially with separation between waveguides for all wavelengths $\lambda$ in our range of interest, and thus, couplings between waveguides further apart than nearest-neighbors are increasingly exponentially suppressed. The lattice in this approximation is shown in Fig. \ref{Fig1}d-f for various values of $L/s.$ The unit cell (marked in green) has six waveguides, and we distinguish two types of coupling terms: those between waveguides within a unit cell, of strength $c_{int}$, and those between waveguides belonging to neighboring unit cells, of strength $c_{ext}$. As we vary the separation $s$ between waveguides within the unit cell while keeping the separation $L$ between unit cells constant, both internal and external coupling terms are modified as $c_{int} =C e^{-\kappa s}$, $c_{ext} =C e^{-\kappa (L-2s)}$, where $C=C(\lambda)$ and $\kappa = \kappa (\lambda)$ are wavelength-dependent experimental parameters. The internal and external coupling terms are represented by black and red lines respectively in Fig. \ref{Fig1}d-f. The color of individual waveguides represents the `chiral charge,' and we see that in this limit the chiral symmetry exists since there are no couplings between waveguides having the same chiral charge. 

As detailed in Section \ref{sec:classification} of the supplementary information, the bulk topology of the tight-binding Hamiltonian for our photonic crystal in the above mentioned approximation, having both chiral and $C_6$ symmetries, is indicated by the topological invariant $[M] \in \mathbb{Z}$, which takes the values
\begin{align}
[M] = \left\{ \begin{array}{c}
0\quad |c_{int}/c_{ext}| > 1\\
2\quad |c_{int}/c_{ext}| < 1
\end{array}\right..
\end{align}
Equivalently, $[M]=2$ for $L/s<3$ and $[M]=0$ for $L/s>3$. This topological invariant is actually protected by $C_2$ symmetry, and would survive even if $C_6$ were broken down to $C_2.$ In the $[M]=0$ phase, our photonic crystal can be adiabatically connected to a photonic crystal having no inter-cell coupling (i.e., the `atomic limit', in an analog atomic crystal) without closing the energy gap. This is the trivial phase. When $[M]=2$, such a connection is not possible, unless we close the bulk energy gap, which amounts to going through a phase transition, or break the symmetry, which is forbidden in this context. Thus, $[M]=2$ signals a different phase when $C_6$ (and hence $C_2$) is present. This new phase has non-trivial topology as the subspace of negative (or positive) energy bands gets inverted or `twists' across the Brillouin zone. Furthermore, it is a crystalline topological phase, as it is protected by $C_2$ symmetry. As we detail below, one observable consequence of the non-trivial topology are energy-degenerate corner-localized topological bound states. 
Furthermore,  chiral symmetry, restricts the values of the bound state energies to be pinned at $\beta=0$ up to corrections that are exponentially small in the system size. 

The stability/protection of the corner modes is captured by an integer index $\mathcal{N}$ as follows. As a whole, $|\mathcal{N}|$ counts the number of stable modes pinned at $\beta=0$ bound to particular defect or corner. It can be expressed as $\mathcal{N}=N_+-N_-$, where $N_\pm$ are integers that count the number of defect-bound states with chiral charge $\pm1$. In our photonic crystal, the trivial phase always has $N_+=N_-=0$ at any type of defect and corner. The non-trivial phase, on the other hand, has $N_+ = N_- = 1$ in each edge unit cell and $N_+ = N_- = 2$ at $2\pi/6$ corners (see Section \ref{sec:chiral_charge} of the supplementary information for detailed explanation and illustration). In both cases, $\mathcal{N}=0$, i.e., there are no stable, topologically protected defect modes. However, at $2\pi/3$ corners, we have $N_+=2$ and $N_-=1$ or $N_-=2$ and $N_+=1$ (these two options alternate at adjacent corners). In this case $|\mathcal{N}|=1$, i.e., there is one stable mode pinned at $\beta=0$ that is localized at each $2\pi/3$ corner. 

The presence or absence of these modes in each phase is shown in Fig. \ref{Fig2}. To determine the existence of the defect modes we calculate the $\beta$ spectrum of the tight-binding models in the hexagon-shaped lattices in both the trivial phase ($L/s=3.53$) and the non-trivial phase ($L/s=2.61$), as shown in Figs. \ref{Fig2}a and \ref{Fig2}b, respectively. Only bulk bands exist in the trivial phase, whereas three types of boundary modes emerge in the non-trivial phase: protected topological zero-modes on the corner unit cells (Fig. \ref{Fig2}c), gapped edge modes (Fig. \ref{Fig2}d), and gapped corner modes (Fig. \ref{Fig2}e). Crucially, although both the topological corner modes and the gapped corner modes are localized together at the corners, they do not hybridize to open a gap. Doing so would require breaking chiral symmetry. Indeed, out of the initial three mid-gap corner modes per corner unit cell, two of them hybridize as chiral-symmetric partners. This results in the gapped corner modes. The topological zero modes, on the other hand, remain at $\beta=0.$ The zero-modes are eigenstates of the chiral operator (see Section \ref{sec:classification} of the supplementary information for details on the operator), with chiral eigenvalues as indicated by the signs in Fig. \ref{Fig2}c.

Let us comment on the stability of these modes in the presence of moderate disorder. 
If the disorder breaks $C_6$ (or even just $C_2$), as it naturally will, but preserves chiral symmetry (e.g., if the disorder is due to small imperfections in the positioning of the waveguides), the topological corner modes can still be localized and pinned at $\beta=0$. Since each topological corner carries a non-vanishing chiral index $\mathcal{N},$ the defect modes remain stable unless they are coupled to another defect mode so that the combined $\mathcal{N}$ vanishes. Modes with opposite chiral indices appear on adjacent corners, so in order to destabilize a defect mode the disorder must be strong enough to generate considerable overlap between modes on different corners.  Hence, if the perturbations are strong enough to close the bulk gap, or to lower the gap in a region such that neighboring corners can easily couple, then the fusion/destabilization of corner modes will occur as they delocalize. Another possibility would be for disorder to nucleate a trivial region with corners inside of a topological region, however, this would only serve to move or distort the location and shape of the protected mode and not destroy it entirely.
If, on the contrary, the perturbations break chiral symmetry while keeping $C_6$ (or just $C_2$) symmetry intact, the corner modes, although they can still remain localized, will generically be lifted away from $\beta=0$ in energy. However, modes on opposite corners will still remain degenerate, and if $C_6$ symmetry is maintained the modes on all six corners will remain degenerate. If the energies of the modes are pushed up to the bulk energies, then light injected at topological corners could couple to other bulk modes with little energy cost, which will lead to the deconfinement of the light. Thus, we conclude that, although both chiral and rotation symmetries are necessary for the confinement of light, this confinement is robust to perturbations that deviate from those ideal symmetric scenarios, and we expect it to be quite robust when chiral symmetry is maintained.  

In what follows we will experimentally probe these modes to conclusively demonstrate the existence of the topological phase.
We use a tunable laser to probe the sample (range 1450 nm to 1650 nm, Keyseight 8164B) and work with fixed sample length of $Z=8$ cm. The control over wavelength of the input beam allows us to control the coupling strengths between the waveguides: the longer the wavelength, the stronger the coupling strengths.  By varying the wavelength we can effectively observe the `dynamics' of the wavefunction as the coupling between waveguides changes.  In the limit of nearest-neighbor coupling, this can be exactly mapped onto time evolution of the wavefunction; but in all cases, observing wavelength response yields a novel and highly useful probe of the mode content and relative propagation constants associated with the wavefunction.  Using this, we will demonstrate that a particular defect mode in the photonic lattice has a topological origin and is pinned to the center of the gap.

To probe the relevant modes, a beam was launched at the input facet of the photonic lattice through a lens-tipped fiber, which allows us to couple the beam into a single waveguide. The difference in the index of refraction between the core of the waveguide and the ambient glass is approximately $\Delta n = 3 \times 10^{-3}$. The radii of the major and minor axes of the waveguides are 4.9 $\mu$m and 3.2 $\mu$m, respectively. In Fig. \ref{Fig3}, we show the diffracted light observed from the output facet for three different wavelengths and in three different cases: trivial phase at $L/s=3.53$ (Fig. \ref{Fig3}a-c), critical phase at $L/s=3.00$ (Fig. \ref{Fig3}d-f), and non-trivial phase at $L/s=2.61$ (Fig. \ref{Fig3}g-i).  Here, we excite the lower-most waveguide which is at the bottom corner of the structure. Increasing the wavelength corresponds to increasing the coupling constant between waveguides, and therefore to an increased rate of diffraction of the optical wavefunction. For the trivial phase, the light injected at a corner simply diffracts into the bulk, since there are no edge states, and therefore no mechanism for confinement to the edge of the structure.  For the critical phase, spreading easily occurs because there is no band gap at all. On the other hand, in the topological phase, light is confined close to the corner at which the light is injected. This confinement occurs across our entire wavelength range, and is associated with the topological corner states that emerge in the topological phase. 

In addition to confinement, the light injected at the bottom corner in the non-trivial phase shows a beating of intensities as a function of wavelength between the modes localized near the corner waveguide. In our experiments, a single waveguide is initially excited and therefore multiple eigenmodes are excited simultaneously since a single waveguide is not an exact eigenstate.  In the case in which the waveguide exactly at the corner is excited (indicated in Fig. \ref{Fig4}a), the two modes that are largely excited are two trivial defect modes at the edge (see Fig. 2e).  The topological mode is actually localized at the two neighbors of the corner waveguide, and is therefore not excited (see Fig. \ref{Fig2}c).  However, when the waveguide just neighboring the corner is excited (indicated in Fig. \ref{Fig4}e), both the topological mid-gap mode and the trivial defect modes and are excited; therefore we see beating between all three.  Since the topological modes are at the center of the band gap, and trivial defect modes have energies symmetrically above and below that of the topological mode, the beating between the trivial defect modes should have precisely double the frequency of that between a trivial defect mode and a topological mid-gap mode.  Therefore, our approach is to measure the beating between the various modes in order to establish that the topological corner mode is at the center of the gap.

We have studied the beating of the eigenmodes by injecting light at a single waveguide at or near the corner. We excite either the lower-most corner waveguide (Fig. \ref{Fig4} first row) or one waveguide away from that corner (Fig. \ref{Fig4} second row), and measure the intensities as a function of wavelength (see supplementary media for animations depicting this beating).
In the first column of Fig. \ref{Fig4}, we show diagrams of the waveguide array at the input facet around the corner of interest, where arrows indicate the waveguide that was initially excited: the lower-most corner waveguide (Fig. \ref{Fig4}a) or its neighboring waveguide to the right (Fig. \ref{Fig4}b). Other columns represent the measurement of light intensity at one of three waveguides, as a function of wavelength: the waveguide to the left of the corner (second column), the waveguide at the corner (third column), and the waveguide to the right of the corner (fourth column), respectively.  We measure and perform a nonlinear least-squares fit on the light intensity with a sinusoidal function  to extract the beating frequencies at each waveguide.

First, with a corner waveguide mode initially excited, we mapped the beating between the trivial defect modes into oscillations of waveguide modes as a function of wavelength. Since in this case only two modes are excited, the oscillation frequency should be the same in every waveguide; this is precisely what we observed, as discussed above.  The frequency was measured to be 12.63 $\mu m^{-1}$, and the ratio of frequencies between the waveguides on and off the corner was $0.98\pm0.09$ (should be 1 ideally). When a neighboring waveguide is excited, the beating between the two trivial defect modes and the one topological mid-gap mode dominates. For this case, the ratio of frequencies between the waveguides on and off the corner was $1.99\pm0.27$ (Fig. \ref{Fig4}f-h) - the predicted value is 2 if the mode is exactly mid-gap.  Indeed, the beating between the trivial defect modes and topological mode can be clearly seen in Figs. \ref{Fig4}f and \ref{Fig4}h, which is half the frequency of the others (measured frequencies for these were 5.95 $\mu m^{-1}$ and 6.50 $\mu m^{-1}$, respectively).  This is the clear signature that the topologically protected mode is precisely at the middle of the band gap.

In this work, we have presented the existence of the zero-dimensional topological defect mode in a 2D time-reversal invariant photonic lattice; this is the first realization of such states in any context, including condensed matter physics, ultracold atoms, or otherwise.  This defect mode is a signature of a new type of crystalline topological phase, has an energy which is topologically protected to lie at mid-gap, and is insensitive to disorder that respects chiral symmetry  (which in this case includes any randomness in the positions of the waveguides that is not large enough to close the band gap near the defect location).  As discussed above, certain perturbations can move the mode from mid-gap (e.g., next-neighbor coupling, which is exponentially suppressed).  That said, our structure provides a limiting case for the modes to {\it start out} at mid-gap rather than at the band edge.  The realization of these modes in photonic crystal slabs and/or photonic crystal fibers could have important technological implications.  For example, resonantly coupling photonic crystal cavity modes is notoriously difficult due to the sensitivity of their resonance frequency to fabrication disorder - if the modes were fully protected, they would necessarily resonantly couple.  In photonic crystal fibers, one important goal is to have small mode volume to enhance nonlinearity.  If modes are guaranteed to be mid-gap, they are necessarily as small as possible, given the gap size.  While it is true that not all fabrication disorder necessarily respects the symmetry required to have rigorous protection, the mode would at the very least {\it start} at mid-gap, rather than sensitively bifurcating from a band edge - giving a topological defect mode a `head start' over other designs. Indeed, protection of zero-modes in two dimensions represents a new phenomenon associated with topological photonic systems and we believe it will have significant implications across a range of optical platforms and devices.   

WAB and TLH are supported by the ONR YIP Award N00014-15-1-2383.  M.C.R. and J.N. acknowledge support from the National Science Foundation under grant number ECCS-1509546; M.C.R. acknowledges support from the Alfred P. Sloan foundation under fellowship number FG-2016-6418.

\newpage
\section{Supplementary information}

\subsection{Topological classification and bulk topological invariants}
\label{sec:classification}

The Bloch Hamiltonian of our photonic crystal in the tight-binding limit with coupling between nearest-neighbor waveguides is
\begin{align}
h({\bf k}) &= c_{ext} h_{ext}({\bf k}) + c_{int} h_{int},
\label{eq:hamiltonian}
\end{align}
where $h_{ext}({\bf k})=\oplus_{i=1}^3 [\cos({\bf k} \cdot {\bf a}_i) \sigma_x + \sin({\bf k} \cdot {\bf a}_i) \sigma_y]$
is due to couplings between waveguides of neighboring unit cells and $h_{int}$ is a matrix with entries $[h_{int}]^{mn}=1$ for nearest-neighbor waveguides $m$, $n$ within the same unit cell, and 0 otherwise.
Here ${\bf a}_1 = (1,0)$, ${\bf a}_{2,3}=(\pm1/2, \sqrt{3}/2)$ are primitive lattice vectors, and the basis of states in the matrices are the six internal degrees of freedom in the unit cell (see Fig. \ref{Fig1}d for numbering). 

The existence of crystalline symmetries expands the topological classification beyond the 10-fold classification \cite{Altland1997} which is built upon time-reversal, particle-hole, and chiral symmetries. In this section we construct the topological classification for crystals in class BDI \cite{Altland1997} with additional $C_6$ symmetry, as these are the symmetries in our tight-binding Hamiltonian \eqref{eq:hamiltonian}. We will then see that our crystalline structure can transition from a non-trivial class to the trivial class as we vary the ratio $s/L$ from $s/L<3$ to $s/L>3$. We begin by pointing out that in BDI class, systems have TR and chiral symmetries
\begin{align}
\hat{T} h({\bf k}) \hat{T}^{-1} &= h(-{\bf k})\nonumber\\
\Pi h({\bf k}) \Pi^{-1} &= - h({\bf k}).
\label{eq:TR_PH_symmetries}
\end{align}
where the TR and chiral operators are $\hat{T} = K$ (where $K$ is complex conjugation) and $\Pi =\sigma_z \oplus -\sigma_z \oplus \sigma_z$. While the TR symmetry is an intrinsic symmetry of photonic systems, chiral symmetry is specific to our lattice structure, and is only approximately preserved (up to exponentially small corrections from further neighbor coupings between the same sublattice in the honeycomb lattice). TR and chiral symmetries imply the existence of (an approximate for the same reason above) particle-hole symmetry 
\begin{align}
\Xi h({\bf k}) \Xi^\dagger = -h(-{\bf k}),
\label{eq:chiral_symmetry}
\end{align}
where $\Xi = \Pi \hat{T}$ is the particle-hole operator. We now consider $C_6$ symmetry,
\begin{align}
\hat{r}_6 h({\bf k}) \hat{r}_6^\dagger = h (R_6 {\bf k}),\;\; \hat{r}_6 = \left(\begin{array}{cccccc}
0&\sigma_0&0\\
0&0&\sigma_0\\
\sigma_x&0&0
\end{array}\right),
\label{eq:rotation_symmetry}
\end{align}
where $\hat{r}_6$ is the rotation operator acting on the internal degrees of freedom of the unit cell, which obeys $[\hat{r}_6, \hat{T}]=0$ and $\hat{r}_6^6=1$, and $R_6$ is the matrix that rotates the crystal momentum by $2\pi/6$ radians. This symmetry implies that the Brillouin zone has the hexagonal shape of Fig. \ref{fig:classification}a. The entire BZ can be generated by rotating the fundamental domain shown by the shaded region in Fig. \ref{fig:classification}a by multiples of $2\pi/6$ rad. In this BZ there are rotation invariant momenta (RIM) ${\bf k}^{(\alpha)}$ which map back to themselves upon a rotation by $\hat{r}_\alpha$ (that is, $\hat{r}_\alpha {\bf k}^{(\alpha)} = {\bf k}^{(\alpha)}$, modulo a reciprocal lattice vector). In $C_6$ symmetric crystals, the RIM are ${\bf k}^{(6)}={\bf \Gamma}$, ${\bf k}^{(3)}={\bf K}$ and ${\bf K'}$ and ${\bf \Pi}^{(2)}={\bf M}$, ${\bf M'}$, and ${\bf M''}$, as seen in Fig. \ref{fig:classification}a. Notice that since ${\bf \Gamma}$ is a 6-fold RIM, it is also a 3-fold and a 2-fold RIM.

The existence of the RIM implies, from \eqref{eq:rotation_symmetry}, that the Hamiltonian commutes with the rotation operator $\hat{r}_\alpha$ at RIM ${\bf k}^{(\alpha)}$,
\begin{align}
[\hat{r}_\alpha,h({\bf k}^{(\alpha)})]=0
\end{align}
Thus, the $\beta$-energy eigenstates at these points of the BZ, $\ket{u^n_{{\bf k}^{(\alpha)}}}$, i.e., the solutions to
\begin{align}
h({\bf k}^{(\alpha)}) \ket{u^n_{{\bf k}^{(\alpha)}}} &= \beta^n({{\bf k}^{(\alpha)}}) \ket{u^n_{{\bf k}^{(\alpha)}}}
\end{align}
are also eigenstates of the rotation operator,
\begin{align}
\hat{r}_\alpha \ket{u^n_{{\bf k}^{(\alpha)}}} &= r^n_\alpha \ket{u^n_{{\bf k}^{(\alpha)}}}.
\end{align}
This allows us to use the rotation eigenvalues $r^n_\alpha$ as labels for the rotation representation of the subspace of negative $\beta$ bands. This is useful since a difference in the group representations of the subspace of negative $\beta$ bands at two m-fold RIM of the BZ implies a non-trivial topology in the system. In particular, we compare the rotation representation at the momenta ${\bf M}$ and ${\bf K}$ with that at ${\bf \Gamma}$, following the construction in reference \cite{Benalcazar2014}, to build topological invariants in $C_6$-symmetric crystals. However, in addition to imposing restrictions on these invariants due to PH symmetry, as in \cite{Benalcazar2014}, we also impose those of TR symmetry. 

Out of all the RIM in the $C_6$-symmetric BZ, we only compare ${\bf M}$ and ${\bf K}$ to ${\bf \Gamma}$ because $C_6$ symmetry identifies the rotation representation in ${\bf K}$ to that in ${\bf K'},$ and the rotation representation in ${\bf M}$ to those in ${\bf M'}$ and ${\bf M''}$, and thus these other points provide redundant topological information. At the 2-fold RIM $\bf M$ we have two rotation eigenvalues $M_1=1$ and $M_2=-1$, while at the 3-fold RIM $\bf K$ we have three rotation eigenvalues $K_1 = 1$, $K_2 = e^{i 2\pi/3}$, and $K_3 = e^{-i 2\pi/3}$ (see Fig. \ref{fig:classification}b). Additionally, at $\bf \Gamma$ we have $\Gamma^{(2)}_1=1$, $\Gamma^{(2)}_1=-1$, as well as $\Gamma^{(3)}_1=1$, $\Gamma^{(3)}_2=e^{i 2\pi/3}$, and $\Gamma^{(3)}_3=e^{-i 2\pi/3}$.
We therefore define the invariants
\begin{align}
[M_i]&=\# M_i - \# \Gamma^{(2)}_i\\
[K_j]&=\# K_j - \# \Gamma^{(3)}_j
\end{align}
for $i=1,2$ and $j=1,2,3$. Here $\# M_i$ is the number of states below the gap in the $\beta$ spectrum that have rotation eigenvalues $M_i$ at RIM ${\bf M}$, and similarly for $\# K_j$, $\# \Gamma^{(2)}_i$, and $\# \Gamma^{(3)}_j$. Out of these five topological invariants, however, some of them are redundant. Since the total number of occupied states is constant over the BZ, we have that
\begin{align}
\# M_1 + \# M_2 &=  \# \Gamma^{(2)}_1 +  \# \Gamma^{(2)}_2\nonumber\\
\# K_1 + \# K_2 + \# K_2 &=  \# \Gamma^{(3)}_1 +  \# \Gamma^{(3)}_2 +  \# \Gamma^{(3)}_3\nonumber
\end{align}
or
\begin{align}
[M_1]+[M_2] = [K_1]+[K_2]+[K_3] = 0
\label{eq:invariant_restrictions_1}
\end{align}
Additionally, TR, PH, and chiral symmetries impose further restrictions on these rotation invariants. Since two of these symmetries imply the third one, we only need to consider restrictions due to two of them. We choose TR and chiral symmetries. The relations between rotation eigenvalues constrained by TR symmetry are due to the fact that the TR and rotation operators commute, $[\hat{r}_\alpha,\hat{T}]=0$, so it follows that
\begin{align}
\hat{r}_\alpha \hat{T} \ket{u^n_{{\bf k}^\alpha}} &= \hat{T} \hat{r}_\alpha  \ket{u^n_{{\bf k}^\alpha}} \nonumber\\
&= \hat{T} r^n_{\bf k^\alpha}  \ket{u^n_{{\bf k}^\alpha}} \nonumber \\
&= (r^n_{\bf k^\alpha})^* \hat{T} \ket{u^n_{{\bf k}^\alpha}},
\end{align}
where the asterisk stands for complex conjugation. Now, if $\ket{u^n_{\bf k}}$ is an eigenstate of $h({\bf k})$ with eigenvalue $\beta_n({\bf k})$, then $\hat{T} \ket{u^n_{\bf k}}$ is an eigenstate of $h(-{\bf k})$ with the same eigenvalue $\beta_n({\bf k})$ [c.f. \eqref{eq:TR_PH_symmetries}]. Thus, more directly we have
\begin{align}
\hat{r}_\alpha \hat{T} \ket{u^n_{{\bf k}^\alpha}} &= r^n_{-\bf k^\alpha} \hat{T} \ket{u^n_{{\bf k}^\alpha}}.
\end{align}
Comparing the last two expresions we conclude that the rotation eigenvalues under TR symmetry obey
\begin{align}
r^n_{-\bf k^\alpha} = (r^m_{\bf k^\alpha})^*\;\; \mbox{for}\;\; \beta^n(-{\bf k}^\alpha) = \beta^m(\bf k^\alpha).
\end{align}
In particular, at time-reversal invariant momenta (TRIM) that are also RIM, $-{\bf k}^\alpha = {\bf k}^\alpha$ (up to a reciprocal lattice vector), if the $\beta$ eigenstates at ${\bf k}^\alpha$ are non-degenerate, the rotation eigenvalues are real, while if they are $\beta$-degenerate the rotation eigenvalues can also come in complex conjugate pairs. In the case of $C_6$-symmetric crystals, $\bf M$, $\bf M'$, and $\bf M''$ are both TRIM and RIM. Since they have eigenvalues of $\pm1$, TR symmetry does not impose restrictions on them. Regarding $\bf K$ and $\bf K'$, since $-{\bf K} = {\bf K'}$, the restriction above reads as
\begin{align}
\# K_1 &= \# K'_1 \nonumber\\
\# K_2 &= \# K'_3 \nonumber\\
\# K_3 &= \# K'_2, \nonumber
\end{align}
which, once added to the condition due to $C_6$ symmetry,
\begin{align}
\# K_j = \# K'_j,\nonumber
\end{align}
for $j=1,2,3$ leads to the relation between invariants,
\begin{align}
[K_2] = [K_3].
\label{eq:invariant_restrictions_2}
\end{align}
So, taking into account the relations between invariants in \eqref{eq:invariant_restrictions_1} and \eqref{eq:invariant_restrictions_2}, we see that only two invariants are necessary, since they determine the value of the remaining three under TR and $C_6$ symmetries. We take this invariants to be
\begin{align}
[M] &= \#M_1 - \# \Gamma^{(2)}_1\\
[K] &= \#K_1 - \# \Gamma^{(3)}_1.
\end{align}
The topological classes in TR invariant crystals with $C_6$ symmetry can then be specified by the two invariants above. The classification thus lies on a two-dimensional vector space specified by the vector
\begin{align}
\chi^{(6)} = ([M],[K]),
\end{align}
for $[M]$, $[K] \in \mathbb{Z}$.

Finally, we impose the constraints on the invariants due to chiral symmetry. Under this symmetry, if $\ket{u^n_{\bf k}}$ is an eigenstate of $h({\bf k})$ with eigenvalue $\beta_n({\bf k})$, then $\Pi \ket{u^n_{\bf k}}$ is an eigenstate of $h({\bf k})$ with eigenvalue $-\beta_n({\bf k})$ [c.f. \eqref{eq:chiral_symmetry}], i.e., $\ket{u^n_{\bf k}}$ and $\Pi \ket{u^n_{\bf k}}$ are partners on opposite sides of the $\beta$ spectrum (having opposite energies). Now let us consider what happens if $[\hat{r}_\alpha, \Pi]=0.$ In this case we have
\begin{align}
\hat{r}_\alpha \Pi \ket{u^n_{\bf k^\alpha}} &= \Pi \hat{r}_\alpha  \ket{u^n_{\bf k^\alpha}}\nonumber\\
&= \Pi r^n_{\bf k^\alpha}  \ket{u^n_{\bf k^\alpha}}\nonumber\\
&= r^n_{\bf k^\alpha} \Pi \ket{u^n_{\bf k^\alpha}}.
\end{align}
Thus, the rotation eigenvalues come in pairs, one on each side of the gap. Now, since $\hat{r}_\alpha$ is a constant operator (i.e. independent of the crystal momentum), its spectrum is the same at any $\alpha$-fold RIM. Thus, the total number of states over both negative and positive $\beta$ bands corresponding to a particular rotation eigenvalue also has to be constant across all the $\alpha$-fold RIM. Thus, if $[\hat{r}_\alpha, \Pi]=0$ we have $2 \# k^{(\alpha)}_i = 2 \#\Gamma^{(\alpha)}_i$, for all $i \in 1,\ldots, \alpha$ which leads to trivial invariants,
\begin{align}
[k^{(\alpha)}_i] = 0\;\; \mbox{if}\;\; [\hat{r}_\alpha, \Pi]=0
\end{align}
for $i \in 1,\ldots, \alpha$. In particular, our model has operators that obey
\begin{align}
[\hat{r}_2,\Pi] \neq 0,\;\;
[\hat{r}_3,\Pi] = 0\nonumber
\end{align}
and we verify that it has $[K]=0$ for all ratios $c_{int}/c_{ext}$. Thus, our structure is topologically characterized by the only invariant $[M]$, which can take integer values. In our model we find
\begin{align}
[M] = \left\{ \begin{array}{c}
0\quad \text{for } |c_{int}/c_{ext}| > 1\\
2\quad \text{for }|c_{int}/c_{ext}| < 1
\end{array}\right..
\end{align}
The transition at $c_{int}/c_{ext}=1$ occurs by closing the bulk $\beta$ gap at the $\bf \Gamma$ point of the BZ. This transition point corresponds to the usual honeycomb lattice, which is well known in the context of graphene to have Dirac cones at $\bf K$ and $\bf K'$. The difference in our formulation resides exclusively in our unit cell definition having six instead of two degrees of freedom (see Fig. \ref{Fig1}d). The $\beta$ bands in our model are shown in Fig. \ref{fig:BZ} for the trivial and non-trivial phases, as well as at the transition point.

\subsection{Weak invariants}
\label{sec:weak_invariants}

In addition to the bulk invariants described above, crystalline systems have two additional weak $Z_2$-valued topological invariants, given by
\begin{align}
\nu_i=\frac{1}{2\pi}\oint_{\mathcal{C}_i} \tr(\A)\;\;\mbox{mod 1},
\label{eq:weak_invariant}
\end{align}
where $\A^{mn}({\bf k})=-i\braket{u_{\bf k}^m}{d u_{\bf k}^n}$ is the Berry connection of negative $\beta$ bands $m$ and $n$, and $\mathcal{C}_i=\pi {\bf b_i} + s \epsilon_{ij} {\bf b_j}$ is a closed path on the boundary of the BZ along the direction of the reciprocal lattice vector $\epsilon_{ij} {\bf b_j}$. These invariants form a $Z_2$-valued reciprocal lattice vector
\begin{align}
{\bf G}_\nu = 2\pi (\nu_1 {\bf b_1} + \nu_2 {\bf b_2})
\end{align}
which indicates the existence of weak topological insulators along the direction ${\bf G}_\nu$. However, in $C_3$ symmetric systems, as this one, this invariant is always zero, as can be seen as follows. The reciprocal lattice unitary vectors ${\bf b_1} =(1,0)$ and ${\bf b_2} =(1/2,\sqrt{3}/2)$ change, upon a $2\pi/3$ rotation, as $R_3{\bf b_1} = {\bf b_2}$ and $R_3{\bf b_2} = -{\bf b_1}-{\bf b_2}$. Now, $C_3$ symmetry requires ${\bf G}_\nu$ to remain invariant under a $C_3$ rotation. Performing this rotation
\begin{align}
R_3{\bf G}_\nu &= 2\pi \left[ \nu_1 {\bf b_2} + \nu_2 (-{\bf b_1}-{\bf b_2})\right]\nonumber\\
&= 2\pi \left[ -\nu_2 {\bf b_1} + (\nu_1-\nu_2) {\bf b_2}\right] \nonumber
\end{align}
we conclude that $\nu_1 = -\nu_2$ and $\nu_2 = \nu_1-\nu_2$ mod 1, or $3 \nu_1=0$ mod 1. Thus, $\nu_1=\nu_2=0$.

\subsection{Zero energy modes: chiral charge and topological protection}
\label{sec:chiral_charge}

A physical consequence of our photonic crystal in the non-trivial phase is the existence of corner-localized modes pinned at zero $\beta$, which are topologically protected only at $2\pi/3$ corners. A topological argument can be made which explains the existence of these modes in the non-trivial phase, and which is easy to picture. Consider Fig. \ref{fig:chiral_charge}. In the non-trivial phase, $c_{int} < c_{ext}$ (see Fig. \ref{fig:chiral_charge}a for a configuration in the non-trivial phase). Even though a physical system will never have $c_{int} = 0$, as this would represent infinitely large unit cells, any system in the non-trivial phase can be adiabatically connected to the system having $c_{int} = 0$ without closing the energy gap. Thus, the crystal in the limit $c_{int}=0$ is also in the non-trivial phase $[M]=2$. In this limiting case, we can read off the numbers $N_\pm$ by counting the number of zero-energy modes per edge or corner unit cell. There is one zero-energy mode at each uncoupled waveguide in Fig. \ref{fig:chiral_charge}b. We see that at edge unit cells we have two zero modes, one of each chirality, (i.e., one `orange' and one `blue'). Thus, $N_+=N_-=1$, and $\mathcal{N}=0$. At $2\pi/6$ corners we have four zero modes, two of each chirality, (i.e., two `orange' and two `blue'). Thus, $N_+=N_-=2$, which also leads to $\mathcal{N}=0$. Finally, at $2\pi/3$ corners, there are three zero modes, two of one chirality and one of the other one (i.e., two `orange' and one `blue' at the upper right corner and  two `blue' and one `orange' at the lower left corner). Thus, $N_+=2$ and $N_-=1$ or viceversa, which results in $|\mathcal{N}|=1$. 

We now turn on $c_{int}$ back to a non-zero value, $c_{int}>0$. These couplings hybridize some of the zero energy modes, spliting their energies away from zero. This energy splitting, however, must conform to the restrictions imposed by chiral symmetry. Concretely, zero energy modes hybridize only in pairs that have canceling total chirality. To see how this is the case, let us consider the basis in which the chiral operator is diagonal, 
\begin{align}
\Pi = \left( \begin{array}{cc}
\mathbb{I}_{3\times 3} & 0\\
0 & -\mathbb{I}_{3\times 3}
\end{array}
\right).
\end{align}
Pictorially, we have assigned the sector with chiral eigenvalue or `chiral charge' of $+1$ ($-1$) to orange (blue) waveguides. In this basis, chiral symmetry \eqref{eq:TR_PH_symmetries} implies that the Hamiltonian has the form
\begin{align}
h({\bf k}) = \left( \begin{array}{cc}
0 & q({\bf k})\\
q^\dagger({\bf k}) & 0
\end{array}
\right) 
\end{align}
where $q({\bf k})$ is a $3 \times 3$ matrix. From the off-diagonal form of the Hamiltonian it follows that there is no coupling between waveguides belonging to the same chiral sector. All couplings exist only between waveguides of opposite chiral sectors. Thus, if initially there are $N_+$ and $N_-$ zero modes, only $N_+$ of them (if $N_+ < N_-$) or  $N_-$ of them (if $N_+ > N_-$) can hybridize once we turn on $c_{int}$, leaving behind $|\mathcal{N}|=|N_+-N_-|$ still pinned at $\beta = 0$.

In our system it follows then that only $2\pi/3$ corners have one robust mode pinned at $\beta=0$, while edges and $2\pi/6$ corners have none. 

To complete the argument, we show what happens in the opposite limiting case. In Fig. \ref{fig:chiral_charge}c the photonic crystal is in the trivial phase. It is adiabatically connected to the crystal shown in Fig. \ref{fig:chiral_charge}d, which has $c_{ext}=0$. Notice that in this limiting case there are no uncoupled waveguides. The eigenmode energies are equally gapped at each unit cell, with no special in-gap states at either edges or corners.

\subsection{Animations}
\label{sec:animations}

\begin{figure*}
\includegraphics[width=160mm]{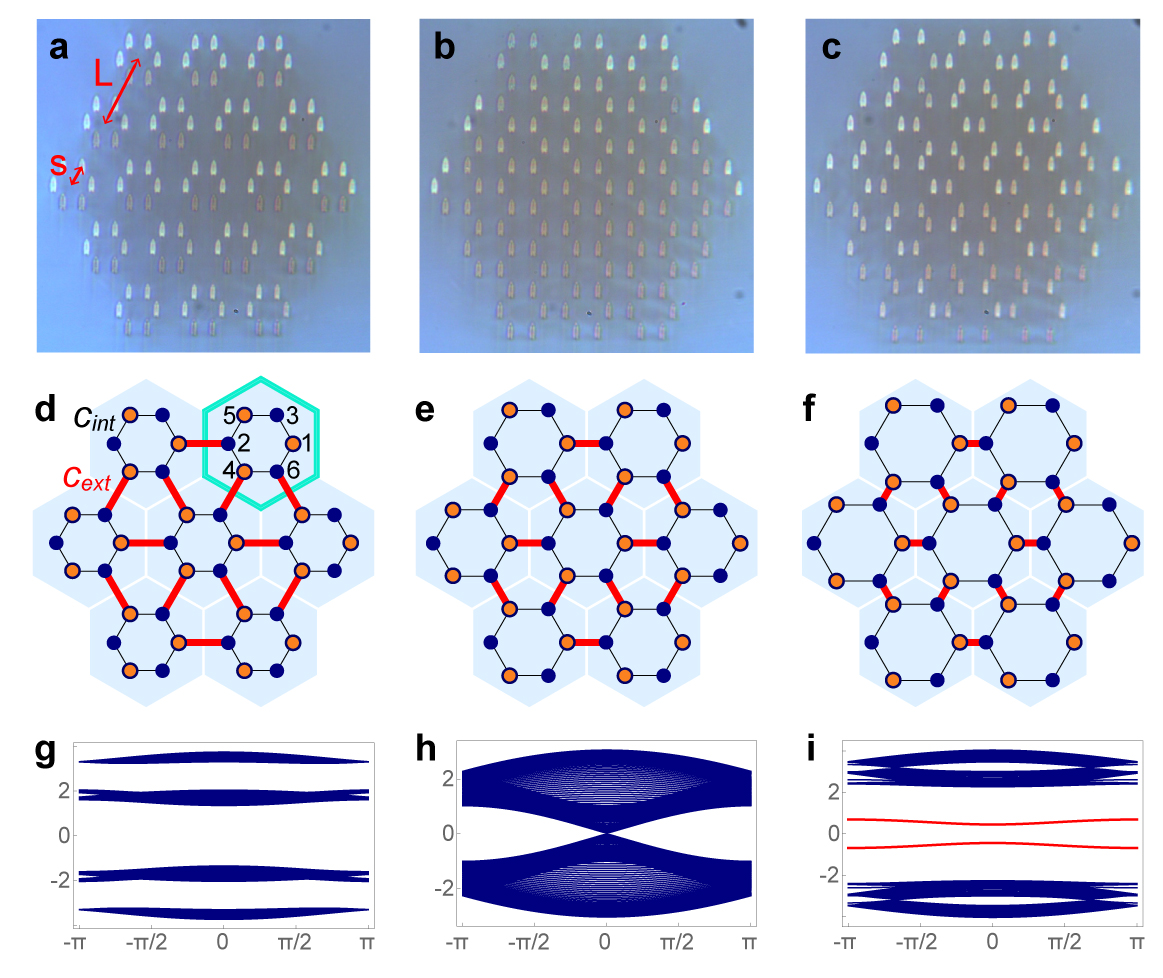}
\caption{\textbf{$C_{6}$ symmetric photonic lattices and band structures}. Each column corresponds to lattices having different $L/s$ ratio, thus belonging to different topological phases: (\textbf{left}) trivial, with $L/s=3.53$, (\textbf{center}) critical at $L/s=3.00$, and (\textbf{right}) non-trivial, with $L/s=2.61$. \textbf{a-c}, Cross-sectional microscope images of the input facet of the photonic waveguide lattices. Light propagates through the structure along the axis perpendicular to the page. \textbf{d-f} Scaled diagrams of the lattices. Green hexagon in \textbf{d} delimits a unit cell.  Black thin (red thick) lines represent intra-cell (extra-cell) couplings of strength $c_{int}$ ($c_{ext}$) in the tight-binding approximation. Color of waveguides represents their chiral charge. \textbf{g-i} Band dispersion calculated using the tight-binding approximation for a configuration with closed boundaries along one direction and open along the other one for crystals with $25$ waveguides along the open direction. The mid-gap bands in \textbf{i} (shown in thick, red lines) have eigenstates localized at edges.}
\label{Fig1}
\end{figure*}

\begin{figure*}
\includegraphics[width=160mm]{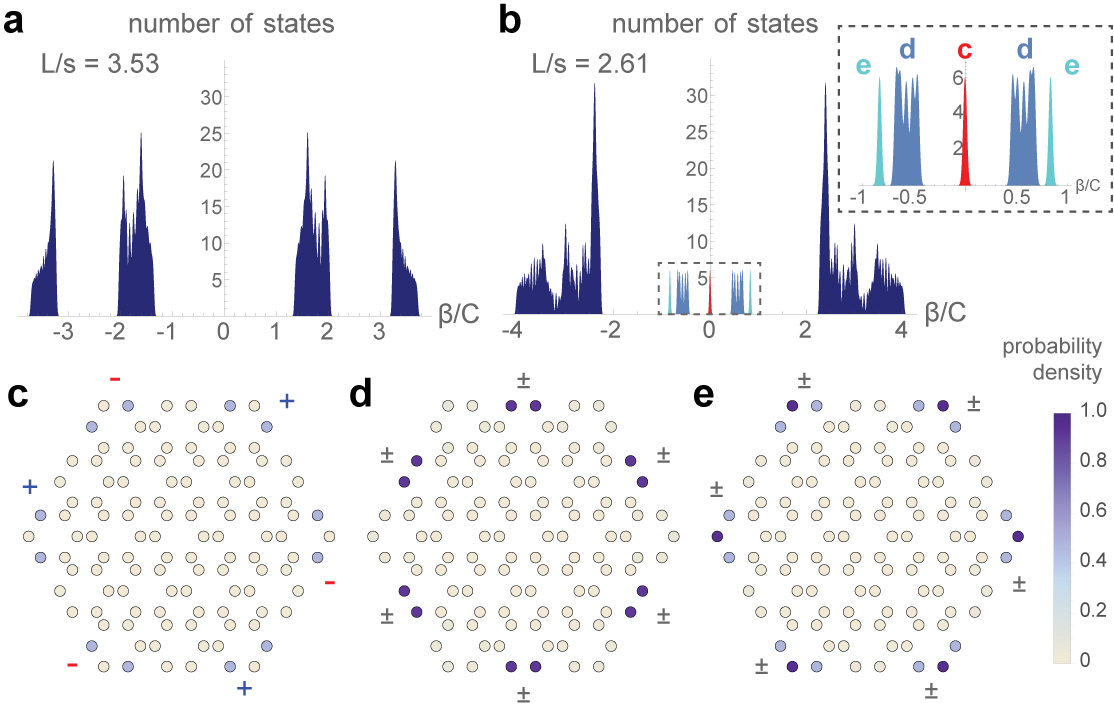}
\caption{\textbf{Numerically calculated density of states (DOS) and eigenmode probability density functions (PDF) of the defect-bound and edge modes using tight-binding approximation for hexagonally shaped lattices (i.e., with full open boundaries). a,} DOS of lattice in the trivial phase ($L/s=3.53$.) \textbf{b,} DOS of lattice in the non-trivial phase ($L/s=2.61$.) Inset: Enlarged DOS around $\beta=0$. We used a larger system size for this numerical calculation (127 unit cells simulated, 19 in experiment) for clearer isolation of the defect-bound modes. Inset labels correspond to PDF indicated in \textbf{c-e}. \textbf{c,} Combined PDF of the six topologically-protected defect-bound modes. \textbf{d,} Combined PDF of the twelve edge modes. \textbf{e,} Combined PDF of the twelve unprotected defect-bound modes. Both the protected and the unprotected defect modes are localized at the \textit{corner unit cells}. However, only unprotected modes occupy the \textit{corner waveguides}. In \textbf{c-e}, the $\pm$ signs indicate the chirality eigenvalues over the subspace spanned by the corresponding edge and corner modes.}
\label{Fig2}
\end{figure*}

\begin{figure*}
\includegraphics[width=160mm]{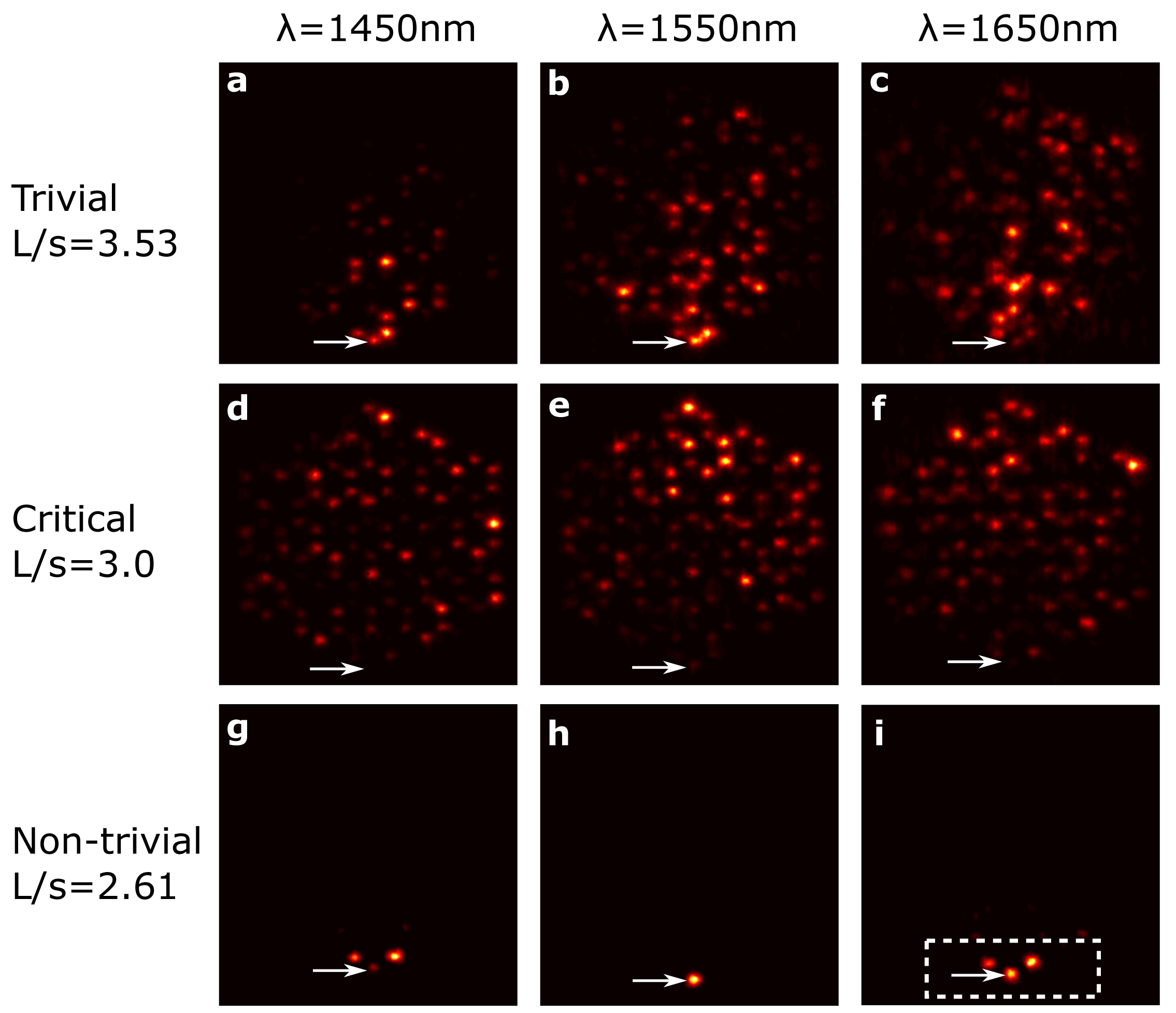}
\caption{\textbf{Experimentally measured evolution of diffracted light at the output facet at different wavelengths. a-c,} Image of the diffracted light in the trivial phase ($L/s=3.53$) measured at the output facet. \textbf{d-f,} Image of the diffracted light at the critical point ($L/s=3.00$). \textbf{g-i,} Image of the diffracted light in the non-trivial phase ($L/s=2.61$). Columns correspond to injection of light with wavelengths $\lambda=$1450 (left column), 1550 (center column) and 1650 nm (right column).  White arrows indicate the position of light injection. In the trivial phase and at the critical point, light increasingly scatters into the bulk as wavelength increases. On the other hand, in the non-trivial phase, light is kept localized near its injection corner within the wavelength range of measurement. In addition, beating of intensities between the corner waveguides enclosed by the white dashed box in \textbf{i} is observed as a function of wavelength (see Fig. \ref{Fig4}).}
\label{Fig3}
\end{figure*}

\begin{figure*}
	\includegraphics[width=160mm]{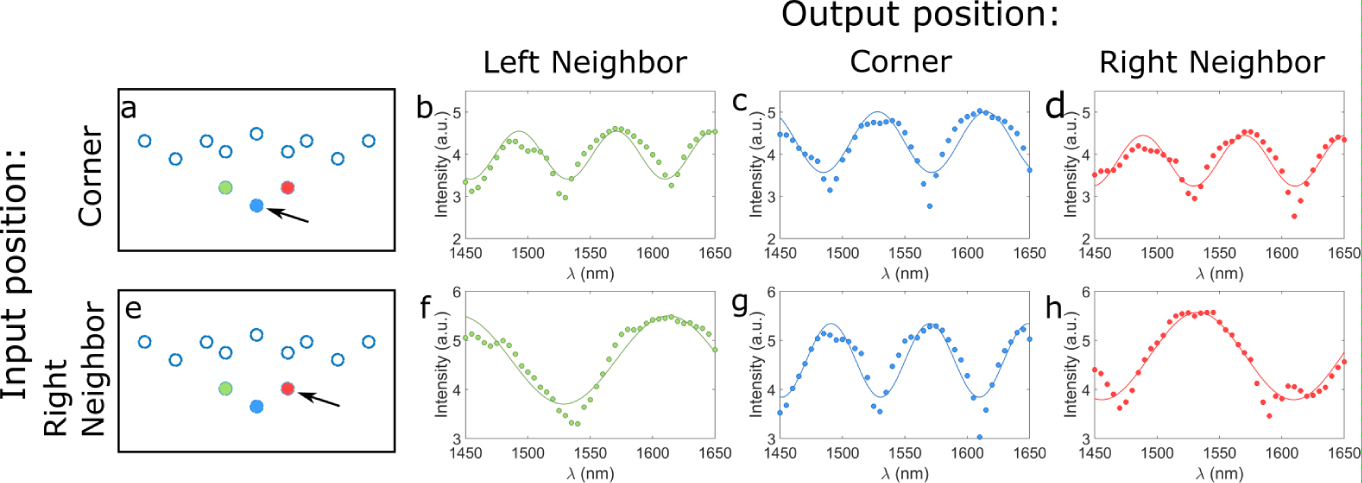}
	\caption{\textbf{Measurements of light intensity at the corner waveguides of the output facet in the non-trivial phase and estimation of its beating frequencies as a function of wavelength. a, } Diagram of the input facet of the waveguide array in non-trivial phase, zoomed-in around the corner where the light is initially injected and localized throughout its propagation (Fig. \ref{Fig3}i). The arrow indicates the waveguide where light was injected for the measurements in (b-d). \textbf{b-d,} Measured light intensities at the output facet at waveguides to the left of the corner (green), at the corner (blue), and to the right of the corner (red), respectively, for light injection as shown in (a). \textbf{e,} Diagram of the input facet of the waveguide array. The arrow indicates the initially excited waveguide for measurements in (f-h). \textbf{f-h,} Measured light intensities at each waveguide on the output facet for light injection as shown in (e).  Solid lines are least-squares fit using a sinusoidal function to measure the beating frequencies. When the light is injected at the corner waveguide, the ratio of beating frequencies between the waveguides on and off the corner is approximately 1, which indicates that only the trivial defect modes are excited. On the other hand, when light is injected at one waveguide away from the corner, the corresponding ratio is approximately 2, which indicates that both trivial and topological defect modes are excited, and this topological mode has $\beta = 0$.}
\label{Fig4}
\end{figure*}

\begin{figure*}[t]
\includegraphics[width=160mm]{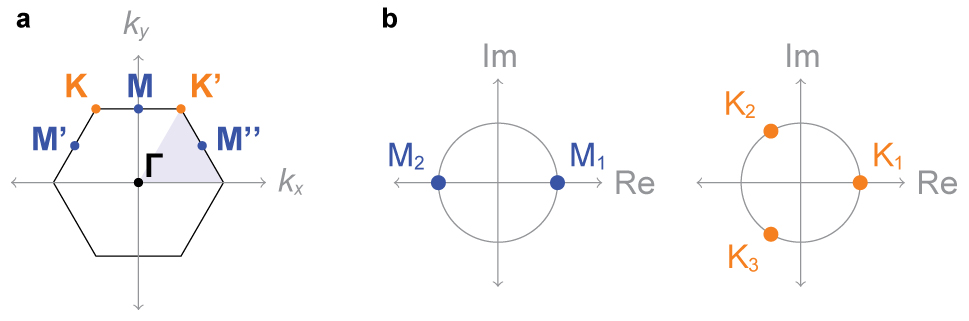}
\caption{\textbf{a} Brillouin zone of the photonic crystals with $C_6$ symmetry and its rotation invariant points. \textbf{b} Unit circle in the complex plane and the rotation eigenvalues at {\bf M} (left) and {\bf K} (right).}
\label{fig:classification}
\end{figure*}

\begin{figure*}[t]
\includegraphics[width=160mm]{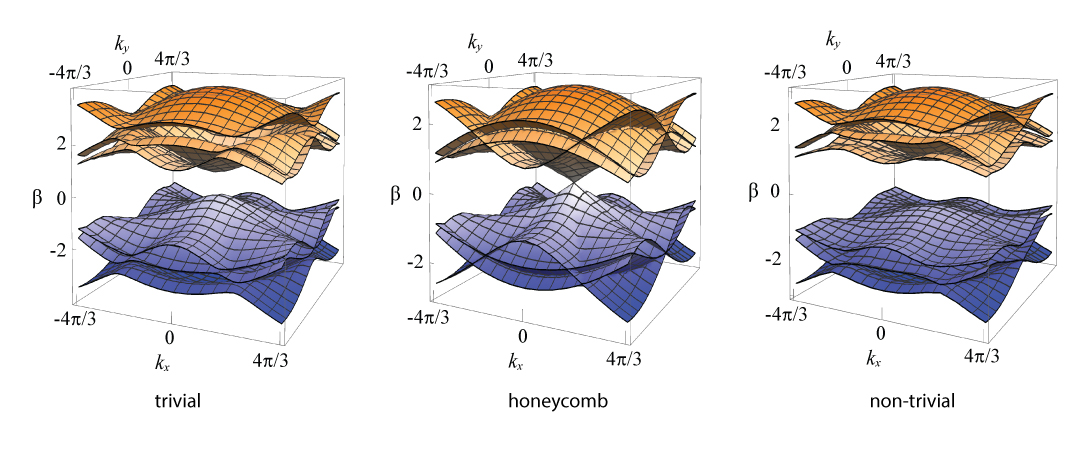}
\caption{Brillouin zones for the crystal in the trivial (left), critical (center), and non-trivial (right) phases.}
\label{fig:BZ}
\end{figure*}

\begin{figure}[t]
\centering
\includegraphics[width=\columnwidth]{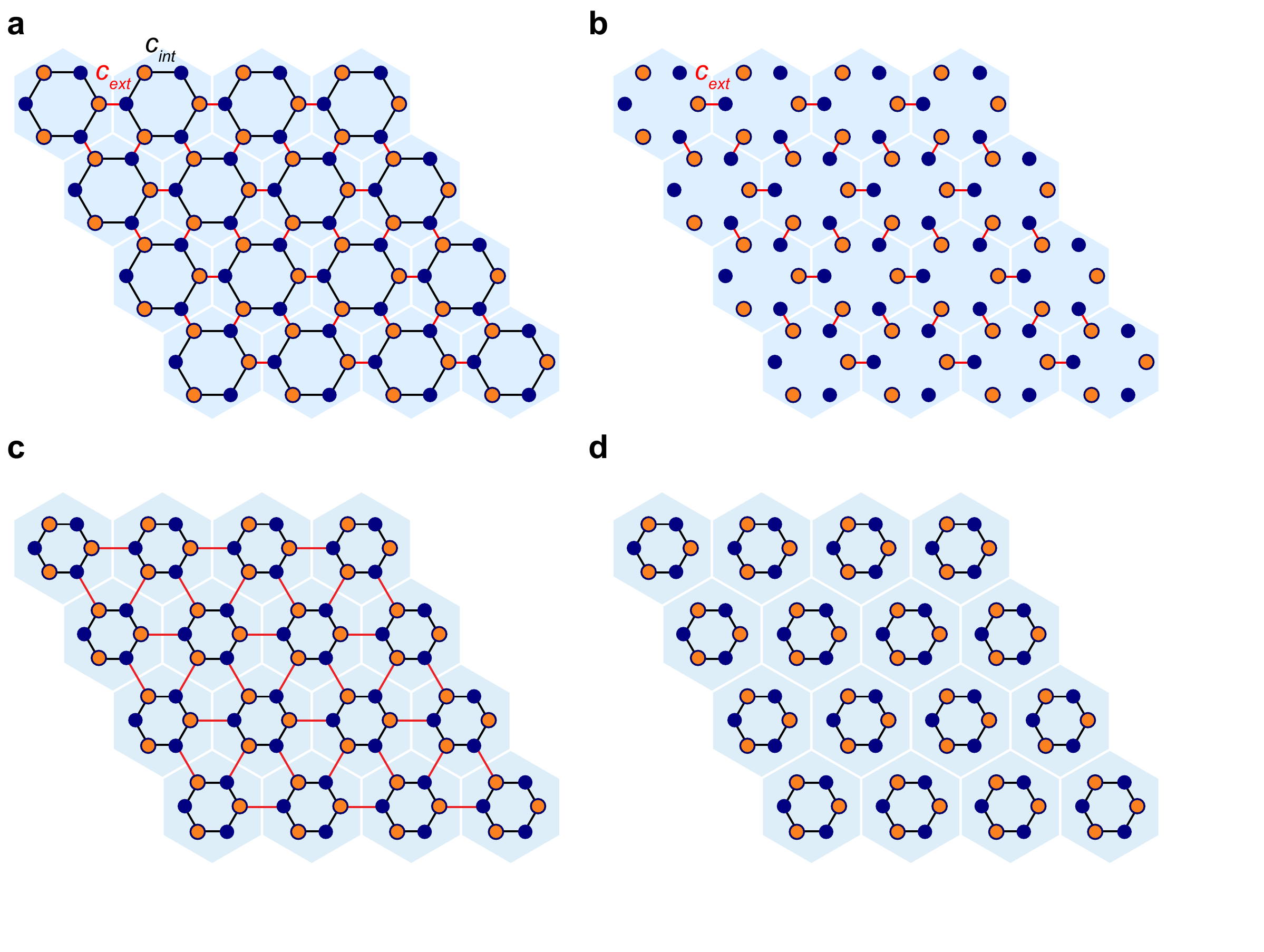}
\caption{\textbf{a} A configuration of our photonic crystal in the non-trivial phase. \textbf{b} Configuration as in \textbf{a} but with $c_{int} = 0$. Each uncoupled waveguide hosts a zero-$\beta$ energy mode. Tight-binding Hamiltonians in both {\bf a} and {\bf b} are in the same non-trivial phase $[M]=2$. \textbf{c} A configuration of our photonic crystal in the trivial phase. \textbf{d} Configuration as in \textbf{c} but with $c_{ext}=0$. Tight-binding Hamiltonians in both {\bf c} and {\bf d} are in the trivial phase $[M]=0$. Orange and blue colors represent chiral charge of $\pm1$ respectively.}
\label{fig:chiral_charge}
\end{figure}

In this supplementary section, we present animations corresponding to the experimental data presented in the text, together with corresponding beam-propagation simulations.
The first animation (Movie1.gif) is an experimental result of optical propagation through $C_{6}$ symmetric photonic lattice with $L/s=2.61$ (corresponding to Fig. \ref{Fig1}c) when the bottom corner waveguide mode was initially excited for a range of wavelengths (Fig. \ref{Fig4} first row).  The oscillation of the light intensity at the output facet is measured in steps of 5 nm from 1450 nm to 1650 nm, which occurs due to the beating between the trivial defect modes. Oscillation frequencies of all three waveguide modes are the same.
The second animation (Movie2.gif) is a similar experimental result when at one waveguide away from the lower-most corner waveguide was initially excited (Fig. \ref{Fig4} second row). The oscillation frequency of the corner waveguide is double the frequency of the other two, which occurs due to the beating between the topological mid-gap mode and the trivial defect modes.
The third animation (Movie3.gif) is a beam-propagation simulation that corresponds to the case of the first animation (Movie1.gif), where the bottom corner waveguide mode was initially excited. Parameters of the simulation are: $\Delta n = 4.5\times 10^{-3}$ and the radii of the major and minor axes are 4.3$\mu$m and 3.6$\mu$m, respectively. The fourth animation (Movie4.gif) is a similar simulation result that corresponds to the case of the second animation (Movie2.gif) with same simulation parameters. These beam-propagation simulations show good agreement with the experimental result. In addition, fifth animation (Movie5.gif) is the beam-propagation simulation result that shows how the beam evolves along the $z$ axis of the sample when the initial beam was incident on the lower-most corner waveguide, with same simulation parameters as above. It shows the characteristic of having same oscillation frequencies for all three waveguide mode as in the first animation (Movie1.gif). The last animation (Movie6.gif) is a similar simulation result of beam propagation along the $z$ axis that corresponds to the case of the second animation (Movie2.gif) with same simulation parameters as above.

\newpage

\newpage

\newpage

\newpage

\newpage
\clearpage
\bibliography{References}

\end{document}